\documentclass[aps,prl,twocolumn,superscriptaddress,amsmath,amssymb,floatfix,preprintnumbers]{revtex4-2}

\usepackage{graphicx}
\usepackage[colorlinks=true,linkcolor=blue,citecolor=blue,urlcolor=blue]{hyperref}

\begin{document}

\preprint{YITP-26-131}

\title{From the Old to the New Starobinsky Inflation}

\author{Shinji Hirano}
\email{shinji.hirano@gmail.com}
\affiliation{School of Science, Huzhou Normal University, Huzhou 313000, Zhejiang, China}
\affiliation{Center for Gravitational Physics and Quantum Information, Yukawa Institute for Theoretical Physics, Kyoto University, Kyoto 606-8502, Japan}

\date{\today}

\begin{abstract}
Starobinsky's anomaly-driven inflation and its \(R+R^2\) (scalaron) formulation are two stages of one evolution: the Euler anomaly supports an early de Sitter phase that decays into the scalaron regime, where observable modes exit. On the exact background, anomaly corrections scale as \(aN_*/(18b)\). The Planck tilt at \(N_*=60\) needs only \(b/a\simeq35\), not the \(b/a\gtrsim10^9\) of early realizations, and the Euler anomaly then lowers the tensor-to-scalar ratio by 17\% at fixed tilt. In the dS analysis, a finite Weyl-coupling range excludes an extra real spin-two pole and keeps the causal response decaying.
\end{abstract}

\maketitle

\textit{Introduction.}---The first model of inflation required no inflaton. Starobinsky observed that the quantum response of many conformal fields, encoded in the trace anomaly, supports a de Sitter (dS) solution of the semiclassical Einstein equations~\cite{Starobinsky:1980te}. The solution is slightly unstable; its slow decay gives a quasi-dS stage that ends in a graceful exit, and its quantum fluctuations produce nearly scale-invariant perturbations~\cite{Mukhanov:1981xt,Starobinsky:1983zz,Kofman:1987ec}. The late stage of this evolution is governed by the local \(R^2\) term in the effective action. Vilenkin's analysis of the full model showed that, once the Hubble rate falls well below its dS value, the anomaly term becomes negligible and the dynamics reduce to those of \(R+R^2\) gravity~\cite{Vilenkin:1985md}, which is conformally equivalent to Einstein gravity with a single scalar degree of freedom, the scalaron~\cite{Whitt:1984pd}. For the grand-unified field content he considered, with a Planckian dS scale, he noted that the observational bounds on the perturbation amplitude require \(b/a\gtrsim10^9\)--\(10^{10}\) in the present notation with the consequence that observable scales leave the horizon in the \(R^2\)-dominated stage; the Weyl anomaly, which does not affect the background, was not included in his treatment of the perturbations. In its \(R+R^2\) form, Starobinsky inflation has become one of the models best supported by cosmic microwave background (CMB) observations~\cite{Planck:2018jri}.

What has remained open is how the full anomaly shapes the observable spectra when this hierarchy is not extreme. In practice the scalaron model is treated as a self-contained theory, with a large \(R^2\) coefficient and an initial state of high curvature taken as inputs, while a treatment of the original anomaly-driven model, in which the anomaly itself shapes the observable modes, must address three issues. First, as we show below, once the observed tilt is imposed, a pivot scale near the anomaly-supported dS solution leads to far too many \(e\)-folds after horizon exit. Second, the Weyl sector of the anomaly, truncated to local form, contains a massive spin-two ghost~\cite{Stelle:1977ry}. Third, when the semiclassical equations are reduced to second order to remove spurious solutions, the anomaly-supported dS phase is lost~\cite{Simon:1991bm,Parker:1993dk}.

In this Letter we extend this picture to the observable spectra, and quantify how the full conformal anomaly modifies the predictions. We treat the anomaly consistently through its Nambu--Goldstone dilaton formulation~\cite{Schwimmer:2010za,Komargodski:2011vj}, which separates the three coefficients \(a\), \(b\) and \(c\) and their distinct roles. The Euler anomaly \(a\) supports the early dS phase; the local \(R^2\) coefficient \(b\) controls its decay into the scalaron regime; and the Weyl anomaly \(c\) modifies the spin-two response. We compute the scalar and tensor spectra on the exact background, without expanding about dS. Scalaron inflation emerges as the limit \(p_+N_*\to0\), where \(p_+\simeq a/(18b)\) is the instability exponent of the dS phase and \(N_*\) the number of \(e\)-folds after the pivot exits the horizon. Dropping the anomaly terms, as in the scalaron formulation~\cite{Linde:2025memorial}, is thus accurate when \(p_+N_*\simeq aN_*/(18b)\ll1\), as for conventional field content, where the dS scale is Planckian and \(a/b\) is tiny. For a large-\(N\) conformal sector with \(a/b\sim10^{-2}\), \(a/b\) is still small, but its effect accumulates over the post-pivot duration, and the deviations are definite and at the \(10\%\) level: at a benchmark reproducing the Planck tilt with \(N_*=60\), the Euler anomaly increases \(1-n_s\) by about \(5\%\) and lowers the tensor-to-scalar ratio \(r_T\) by about \(10\%\), or about \(17\%\) at fixed \(n_s\), both shifts scaling as \(p_+N_*\). The pivot lies about \(60\) \(e\)-folds before the end of inflation, in the second stage, which is why the scalaron description works; the first stage supplies what that description takes as input.

\textit{From the anomaly-supported branch to the scalaron regime.}---We write the physical metric as \(\widehat g_{\mu\nu}=e^{-2\tau}g_{\mu\nu}\), with \(\tau\) the dilaton. The effective action consists of the Einstein--Hilbert term, the local \(R^2\) term, the local Weyl term, and the Wess--Zumino functional of the anomaly:
\begin{align}
S={}&\int d^4x\,\Big\{\frac{M_{\rm Pl}^2}{2}\sqrt{-\widehat g}\,\widehat R+\frac{\sqrt{-g}}{(4\pi)^2}\Big[a\,\mathcal L_{\rm WZ}\nonumber\\
&+b\big(R+6\Box\tau-6(\nabla\tau)^2\big)^2+(\kappa_W-c\,\tau)W^2\Big]\Big\},
\label{eq:action}
\end{align}
where \(\mathcal L_{\rm WZ}=\tau E_4+4G^{\mu\nu}\partial_\mu\tau\partial_\nu\tau-4(\nabla\tau)^2\Box\tau+2(\nabla\tau)^4\) is the Euler Wess--Zumino density, \(E_4\) and \(W^2\) are the Euler density and the square of the Weyl tensor, and the remaining conformal-field information resides in nonlocal functionals of the fiducial metric \(g\)~\cite{Hirano:companion}. We take \(c=a\), as for holographic conformal field theories at large \(N\), and no cosmological constant. This description applies while the matter sector is approximately conformal, between an infrared mass gap \(m_{\rm gap}\) and the ultraviolet cutoff \(\Lambda_{\rm EFT}\) of the four-dimensional effective theory, set by the lowest omitted microscopic threshold or by the breakdown of the four-dimensional description. We assume \(m_{\rm gap}\ll H\) throughout inflation and \(H_0\ll\Lambda_{\rm EFT}\).

Equation~\eqref{eq:action} is local. On the Friedmann--Lema\^itre--Robertson--Walker background the fiducial metric \(g\) is flat, and in the conformal vacuum the effective action of the conformal sector reduces to the anomaly-induced terms together with the local \(R^2\) term, whose coefficient \(b\) is not fixed by the anomaly; since \(W^2=0\), the \(c\) and \(\kappa_W\) terms do not contribute. No nonlocal contribution arises, and the background evolution below is exact for given \(a\) and \(b\). At quadratic order in perturbations, the remaining nonlocal part of the effective action enters only through the Weyl curvature of \(g\), via a Weyl form factor \(F_W\) acting on spin-two modes. We treat it in the local approximation, replacing \(F_W\) by its value \(F_0\) at the on-shell point of the massless graviton (\(z=0\) below), except in the spin-two stability analysis, where the complete dS form factor is used. The matching, its domain of validity, and the spurious massive mode that the local truncation introduces are analyzed in Ref.~\cite{Hirano:companion} (Secs.~2.2 and 6.4, and Appendix~B).

For \(b=0\) the homogeneous equations admit the dS solution
\begin{equation}
H_0^2=\frac{8\pi^2M_{\rm Pl}^2}{a},
\end{equation}
supported by the Euler anomaly alone. The \(R^2\) term renders it weakly unstable: perturbations grow as \(e^{p_+N}\), with
\begin{equation}
\frac{a}{b}=6p_+(p_++3),\qquad p_+\simeq\frac{a}{18b}.
\end{equation}
The background equations are integrated numerically without approximation, and all results below use this exact solution. Its structure is captured by an analytic relation, first obtained in the slow-roll regime~\cite{Vilenkin:1985md,Kofman:1987ec}; at leading order in \(p_+\) and in slow roll, but for any value of \(\epsilon_H/p_+\) (Appendix~A),
\begin{equation}
\frac{H_0^2}{H^2}\simeq1+\frac{2\epsilon_H}{p_+},\qquad
\frac{d\ln\epsilon_H}{dN}\simeq p_++2\epsilon_H ,
\label{eq:closedform}
\end{equation}
where \(\epsilon_H=-\dot H/H^2\) and \(N=\ln\bar a\) is the number of \(e\)-folds, increasing with time and distinct from the post-pivot duration \(N_*\). Its relative corrections are \(\epsilon_H/6\) at next order, so along the benchmark trajectory below it is accurate to about \(10^{-3}\) at the pivot. Equation~\eqref{eq:closedform} is not used in the numerical computation; it serves to organize and explain the results.

\begin{figure*}[t]
\centering
\includegraphics[width=0.8\textwidth]{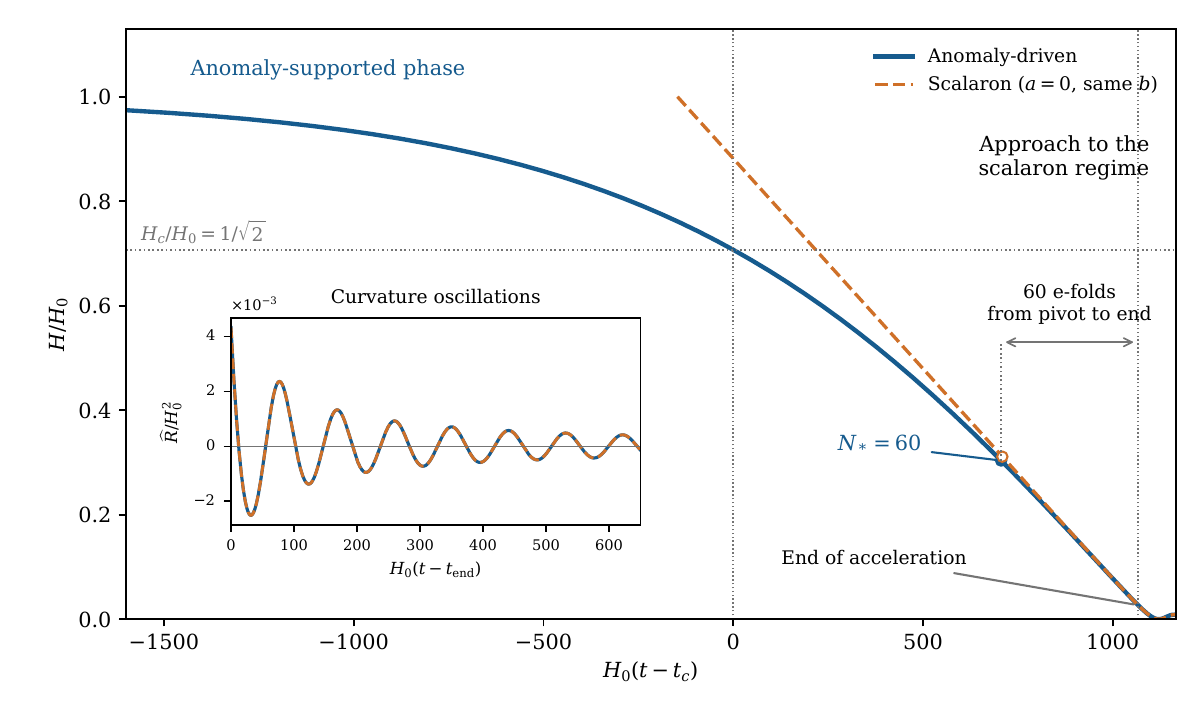}
\caption{Evolution from anomaly-driven inflation to the scalaron regime
for $b/a=34.5$. The solid blue curve is the anomaly-driven solution;
the dashed orange curve is the independent $R+R^2$ solution with the
same $b$ and $G$. Their cosmic-time origins are shifted so that the
first end of acceleration, $\epsilon_H=1$, coincides; they are not
aligned at $t_c$. Here $H_0$ is the anomaly-supported de Sitter scale,
and $t_c$ denotes the anomaly trajectory's crossing of
$H_c=H_0/\sqrt{2}$. The scalaron curve is displayed only for
$H\leq H_0$; this cutoff is not a maximum Hubble rate or a physical
onset of scalaron inflation. The filled blue and open orange circles
mark the anomaly-driven and scalaron pivots, respectively, each
$N_*=60$ $e$-folds before its own end of acceleration. The horizontal
arrow indicates this interval for the anomaly-driven solution.
The inset shows the physical Ricci scalar $\widehat R$, with time
measured from each trajectory's end of acceleration. The two sets
of oscillations are indistinguishable at the plotted resolution,
illustrating the suppression of the Euler-anomaly contribution
in the low-curvature regime.}
\label{fig:evolution_comparison}
\end{figure*}

Without the \(R^2\) term, the first-order equation governing the anomaly branch is singular at \(H_c=H_0/\sqrt2\), where the coefficient of \(\dot H\) vanishes. For \(b\neq0\) this point is crossed regularly. Below \(H_c\) the Einstein term dominates the support, and the evolution approaches the scalaron regime,
\begin{equation}
\dot H\simeq-\frac{M_s^2}{6},\qquad M_s^2=\frac{4\pi^2M_{\rm Pl}^2}{3b},
\end{equation}
which ends at \(\epsilon_H=1\) and is followed by oscillations of the curvature at frequency \(M_s\), with matter-like average expansion. Figure~\ref{fig:evolution_comparison} shows the full evolution for the benchmark discussed below, together with the scalaron solution with the same \(b\) and \(a=0\). The two trajectories approach one another closely before the observable modes leave the horizon; their residual difference, of order \(p_+N_*\), is what the spectra below measure. Their post-inflationary oscillations are indistinguishable.

Combining Eq.~\eqref{eq:closedform} with the slow-roll relation of the scalaron stage, the anomaly's share of the support at the pivot is
\begin{equation}
\frac{H_*^2}{H_0^2}\simeq p_+N_*\simeq\frac{aN_*}{18b}.
\label{eq:share}
\end{equation}
When \(p_+N_*\ll1\), the observable modes leave the horizon deep in the scalaron stage, and the anomaly-supported phase lies before the observable window.

\textit{Primordial perturbations and the CMB pivot.}---We fix the split-Weyl redundancy by \(\delta\tau=0\) and \(\bar\tau=-\ln\bar a\), so that the fiducial metric is Minkowski plus perturbations, and work in longitudinal gauge. With the Weyl sector switched off, the scalar sector then contains a single propagating mode \(\beta_k\), with the curvature perturbation \(\zeta_k=\beta_k/\epsilon_H\), since the other scalar is auxiliary once the background equations hold; the Weyl sector is restored afterwards as a small correction, while the \(a\)- and \(b\)-dependent background is kept exact, so the anomaly-supported phase is retained (Appendix~B). We derive the scalar and tensor quadratic actions on a general Friedmann--Lema\^itre--Robertson--Walker background, evaluate them on the exact numerical solution, and solve the mode equations with Bunch--Davies initial conditions. The pivot \(k_*\) exits the horizon \(N_*\) \(e\)-folds before the end of acceleration. Two independent derivations of the quadratic actions agree through second order in the displacement from dS, the numerical results are converged to a few parts in \(10^5\), and in the limit \(a\to0\) they reproduce an independent Einstein-frame computation of \(R+R^2\) inflation. The method and checks are summarized in the End Matter (Appendices~B and C); further details are given in the companion paper~\cite{Hirano:companion}. All numerical values quoted below are obtained on this exact background; only the Weyl contributions are treated approximately, as described below. The convergence quoted above is a numerical uncertainty; the Weyl-sector treatment carries a separate theoretical uncertainty, estimated in Appendix~B. The analytic relations, such as Eqs.~\eqref{eq:closedform}, \eqref{eq:share} and those below, are leading-order approximations that reproduce these values to the accuracy stated and explain their origin.

The exact scalar spectrum is reproduced by
\begin{align}
A_s&\simeq\frac{1}{72\,b\,\epsilon_{H*}^2},\label{eq:As}\\
n_s-1&\simeq-2\left.\frac{d\ln\epsilon_H}{dN}\right|_*=-2p_+-4\epsilon_{H*},
\label{eq:ns}
\end{align}
where the second equality uses Eq.~\eqref{eq:closedform} and holds at the same order; the tilt itself is computed directly from the exact spectrum. These relations were first derived near dS, but they are slow-roll relations: they depend only on \(p_+\) and \(\epsilon_{H*}\), both small in the slow-roll regime where the observable modes exit, and not on the distance from dS. The exact computation confirms them to about \(2\times10^{-4}\) in \(n_s\), even where \(\epsilon_{H*}/p_+\) is large.

Equation~\eqref{eq:ns} explains both the difficulty and its resolution. For a pivot near dS, \(\epsilon_{H*}\ll p_+\), the post-pivot duration \(N_*\simeq p_+^{-1}\ln[1+p_+/(2\epsilon_{H*})]\) grows without bound as the pivot approaches dS; for the examples we examined, with \(\epsilon_{H*}/p_+=0.1\), matching the observed tilt fixes \(p_+\simeq0.015\) and gives \(120\)--\(170\) \(e\)-folds, well above the \(50\)--\(60\) allowed by standard reheating~\cite{Liddle:2003as}. At fixed \(N_*\), Eq.~\eqref{eq:closedform} instead determines \(\epsilon_{H*}\), and the observed tilt is reached with the pivot in the scalaron stage. As \(p_+\to0\), Eq.~\eqref{eq:ns} reduces to \(n_s-1\simeq-4\epsilon_{H*}\simeq-2/N_*\), the scalaron result. For \(p_+>0\) the spectrum is always redder, and
\begin{equation}
n_s<1-\frac{2}{N_*}
\end{equation}
throughout the range we have examined.

With \(N_*=60\), the Planck central value \(n_s=0.9649\)~\cite{Planck:2018jri} is reproduced for \(b/a=34.5\), at which \(p_+=1.609\times10^{-3}\), \(\epsilon_{H*} = 7.915\times10^{-3}\), \(H_*/H_0=0.3036\) and \(p_+N_*=0.0966\). The amplitude \(A_s=2.1\times10^{-9}\) fixes \(a\simeq3.1\times10^9\), \(b\simeq1.1\times10^{11}\), and \(H_0\simeq3.9\times10^{14}\,\)GeV. Table~\ref{tab:benchmark} compares the predictions with the exact scalaron results, at the same \(N_*\) and at the same \(n_s\).

\begin{table}[t]
\caption{Predictions of the anomaly-driven benchmark (\(b/a=34.5\), \(N_*=60\), \(F_0(H_0)=-7c\); with \(F_0=0\), \(r_T=2.94\times10^{-3}\)) and of scalaron inflation computed exactly, at \(N_*=60\) and at the same \(n_s\) (\(N_*=57.2\)).}
\label{tab:benchmark}
\begin{ruledtabular}
\begin{tabular}{lccc}
& anomaly & \multicolumn{2}{c}{scalaron} \\
& & \(N_*=60\) & same \(n_s\) \\
\hline
\(n_s\) & \(0.9649\) & \(0.9665\) & \(0.9649\) \\
\(r_T\) & \(2.86\times10^{-3}\) & \(3.23\times10^{-3}\) & \(3.55\times10^{-3}\) \\
\(n_T\) & \(-3.6\times10^{-4}\) & \(-4.2\times10^{-4}\) & \(-4.6\times10^{-4}\) \\
\end{tabular}
\end{ruledtabular}
\end{table}

Across \(b/a=34.5\) to \(1000\), the deviations from scalaron inflation at the same \(N_*\) are reproduced by (Appendix~C)
\begin{equation}
n_s\simeq n_s^{\rm scal}-p_+,\qquad
\frac{r_T}{r_T^{\rm scal}}\simeq1-p_+N_*,
\label{eq:deviations}
\end{equation}
with coefficients consistent with \(-1\) across the scan; the \(r_T\) relation refers to the Euler anomaly alone (\(F_0=0\)). The Euler anomaly thus lowers \(r_T\) by \(9\%\) at the benchmark, or \(17\%\) at fixed \(n_s\).

For tensors, the Weyl sector enters through the matched coupling \(F_0\), of the same normalization as \(\kappa_W\), with the dS light-branch normalization evaluated at the pivot, and the tensor-to-scalar ratio is
\begin{equation}
r_T\simeq4R_W\,\epsilon_{H*}^2,\qquad R_W=\frac{72b}{6b-F_0}.
\end{equation}
The truncation-induced massive spin-two pole of this local description, which has negative residue, is absent from the complete dS form factor, written as a function of \(z\equiv(\lambda_0-\lambda)/H^2\), where \(\lambda\) is the eigenvalue of the Lichnerowicz-type operator on transverse-traceless modes and \(\lambda_0=6H^2\):
\begin{equation}
F_W(z)=F_0+c\left[\psi(2)-\psi\!\left(\tfrac12+\sqrt{\tfrac94-z}\right)\right],
\end{equation}
where \(\psi\) is the digamma function. The massless graviton has \(z=0\), where \(F_W=F_0\), and a spin-two mode of mass \(m\) has \(z=m^2/H^2\). The branch cut starting at \(z=9/4\) replaces the real pole by a complex pair \(z_*,\bar z_*\), on the physical sheet. With the retarded prescription, whose Green function contains both late-time exponents of the pair, the resulting causal response decays under Hubble friction provided \(|{\rm Im}\sqrt{z_*-9/4}|<3/2\). With \(\alpha\equiv-F_0/c\), this condition bounds \(\alpha\) from below, while for \(\alpha>\alpha_{\rm crit}\simeq4B+1+2\ln2\), with \(B=3bs/c\) and \(s\simeq4\), a real spin-two pole reappears below the continuum. A finite window therefore exists for the total effective coupling \(F_0\), including any local \(\kappa_W\) contribution. It requires \(F_0=\mathcal O(c)\): if \(F_0\) stayed finite as \(c\to0\), the form factor would reduce to its local truncation, with a real massive pole of negative residue. These two properties, the absence of an extra real pole and a decaying retarded response, are what we establish; they do not by themselves amount to a general proof of quantum ghost freedom. Including the on-shell running \(dF_0/d\ln H=-c\), stability throughout the quasi-dS regime of the benchmark (\(\epsilon_H\lesssim0.1\)), where the criterion is controlled, requires \(F_0(H_0)\lesssim-5c\); we adopt \(F_0(H_0)=-7c\), which leaves a comfortable margin there (Appendix~D), consistent with matching the Weyl coupling at \(m_{\rm gap}\), which gives \(\alpha\simeq\ln(H/m_{\rm gap})+\mathcal O(1)\), for \(m_{\rm gap}\sim10^{-3}H_0\). For scalars, the Weyl correction is evaluated perturbatively and changes \(n_s\) by less than \(10^{-5}\): the light scalar is, to leading order, a conformal fluctuation of the metric with vanishing Weyl curvature, so its contribution is negligible for the form factor and range of couplings considered here. It lowers \(r_T\) by about \(3\%\) at the benchmark. The tensor tilt satisfies the single-field consistency relation \(n_T\simeq-r_T/8\) to about \(2\%\).

\textit{Discussion.}---Within anomaly-driven inflation, scalaron inflation is not an independent ingredient but the late stage of the evolution. Its accuracy as a description of the CMB does not follow from the dynamics that make inflation happen: it follows from where the pivot falls. The pivot lies \(50\)--\(60\) \(e\)-folds before the end of inflation, in the scalaron stage whenever \(p_+N_*\ll1\). This is the regime of Vilenkin's analysis; the present results extend it to a less extreme coefficient hierarchy, for which \(p_+N_*\) produces measurable corrections, and to the observable spectra. The anomaly-supported phase therefore lies outside the observed window, although its residual contribution, of order \(p_+N_*\), is imprinted on the spectra. What it provides is the initial state from which the scalaron stage begins. The large \(R^2\) coefficient, fitted here to the scalar amplitude, can be attributed to the same conformal sector as \(a\) and \(c\), although its magnitude is not derived here.

The benchmark \(N_*=60\) reproduces the CMB-only tilt but requires a stiff post-inflationary phase: instant reheating gives \(N_*\simeq58\), and minimal \(R+R^2\) reheating \(N_*\simeq53\), where the Planck tilt holds within \(1\sigma\) for \(b/a\gtrsim35\)~\cite{Hirano:companion}. Like scalaron inflation, it is in tension at the \(2.5\)--\(3\sigma\) level with the higher tilt preferred by combining CMB and DESI baryon acoustic oscillation data within \(\Lambda\)CDM~\cite{ACT:2025fju,DESI:2025zgx}. With dynamical dark energy the picture changes: for Planck and DESI data in \(w_0w_a\)CDM with broad uniform priors, the inferred tilt is consistent with the Planck-only value~\cite{DESI-ns:2025}, so the model's standing on \(n_s\) is tied to the dark-energy question. The distinctive prediction is a tensor-to-scalar ratio below the scalaron value at fixed \(n_s\): by about \(17\%\) from the Euler anomaly, scaling as \(p_+N_*\), and about \(20\%\) including the Weyl sector, a difference \(\Delta r_T\simeq7\times10^{-4}\). 

The results assume a constant \(b\): in the covariant realization discussed in Ref.~\cite{Hirano:companion}, a curvature-dependent \(b\) carries energy that can overwhelm the anomaly support, and a constant \(b\) requires moduli that change it by well below \(10^{-3}\) during inflation. The normalization requires a conformal sector with \(a\sim10^9\). Apart from the stability analysis, the Weyl-sector results rely on the local approximation to the form factor; the background and the \(F_0=0\) scalar spectrum do not, and the small scalar Weyl correction is evaluated within it. Within these conditions, the old and the new Starobinsky inflation are one theory: the conformal anomaly supports the initial state, and the scalaron governs the observable universe.

\begin{acknowledgments}
The author is grateful to Fl\'avio Coelho, Carlos Herdeiro, and Yuki Sato for valuable discussions and earlier collaborations on related topics that helped motivate the present work, and to Robert de Mello Koch for helpful discussions. He also thanks the Department of Mathematics at Nagoya University for its hospitality during several visits, when part of this work was completed.
The author used ChatGPT (OpenAI; GPT-6 Astra and Sol) and Claude (Anthropic; Claude Opus 5.5) to assist with manuscript preparation and with the analytical and numerical calculations, including writing code; ChatGPT was also used to prepare Fig.~\ref{fig:evolution_comparison}. These tools were used under the author's direction, and their outputs were verified by the cross-checks described in Appendix~B, which the author examined and, for the Mathematica implementations, ran. The author takes full responsibility for the scientific content and conclusions.
\end{acknowledgments}

\section*{End Matter}

\section*{Appendix A: Derivation of Eq.~(\ref{eq:closedform})}
At leading order, Eq.~\eqref{eq:closedform} is equivalent to Vilenkin's slow-roll equation \(H^2-H_0^2=6(H_0^2/M_s^2)\dot H\)~\cite{Vilenkin:1985md}. Here we derive it from an exact identity, which also yields its first correction.
In units \(H_0=1\), with \(h=H/H_0\) and \(\beta=b/a\), the Friedmann constraint of the action~\eqref{eq:action} reads
\begin{equation}
h^2-h^4+6\beta\left(2h\ddot h-\dot h^2+6h^2\dot h\right)=0 .
\label{eq:friedmann}
\end{equation}
With \(\eta_N\equiv d\ln\epsilon_H/dN\), one has \(\dot h=-\epsilon_Hh^2\) and \(\ddot h=-\epsilon_Hh^3(\eta_N-2\epsilon_H)\), and Eq.~\eqref{eq:friedmann} becomes, without approximation,
\begin{equation}
\frac{H_0^2}{H^2}-1=\frac{\epsilon_H\,(6+2\eta_N-3\epsilon_H)}{p_+(p_++3)} ,
\label{eq:identity}
\end{equation}
using \(6\beta=1/[p_+(p_++3)]\). On the numerical background it holds to \(10^{-11}\) or better up to the end of acceleration. When \(\epsilon_H\) and \(\eta_N\) are small, the 6 dominates the bracket, and \(6/[p_+(p_++3)]\simeq2/p_+\) gives the first relation in Eq.~\eqref{eq:closedform}; no assumption on \(\epsilon_H/p_+\) enters. Differentiating it with \(d\ln H/dN=-\epsilon_H\) gives \(\eta_N=p_+H_0^2/H^2=p_++2\epsilon_H\), the second relation. Inserting \(\eta_N\) back into Eq.~\eqref{eq:identity},
\begin{equation}
\frac{H_0^2}{H^2}-1\simeq\frac{2\epsilon_H}{p_+}\left(1+\frac{\epsilon_H}{6}\right),
\end{equation}
where the \(\mathcal O(p_+)\) corrections cancel exactly. The leading form is therefore accurate to \(\epsilon_H/6\): \(1.2\times10^{-3}\) at the benchmark pivot, and \(8\times10^{-6}\) with the correction included. For \(\epsilon_H\ll p_+\) it reduces to the linearized instability; for \(\epsilon_H\gg p_+\), with \(M_s^2=3p_+H_0^2\) at leading order, to \(\dot H=-M_s^2/6\).

\section*{Appendix B: Quadratic actions, numerics and checks}
With \(\delta\tau=0\) and \(g={\rm diag}(-(1+2A),(1-2\psi)\delta_{ij})\), all sectors of Eq.~\eqref{eq:action} are expanded to second order for an unspecified scale factor \(\bar a(\eta)\) and averaged over one wavelength of a real Fourier mode. Because \(\sqrt{-g}\,W^2\) is Weyl invariant, the Weyl terms are independent of \(\bar a\): \(L_W=F_0k^4(A_k+\psi_k)^2/(24\pi^2)\) for scalars and \(L_W=F_0[(h''-k^2h)^2-4k^2h'^2]/(32\pi^2)\) for one tensor polarization. For scalars we use
\begin{equation}
\beta_k=A_k+\frac{\psi_k'}{\mathcal H}+\frac{k^2\psi_k}{3\mathcal H^2}+\epsilon_H\psi_k ,
\end{equation}
with \(\mathcal H=\bar a'/\bar a\). At \(F_0=0\), the coefficient of \(\beta_k''\psi_k\) vanishes identically, and that of \(\psi_k'^2\) is proportional to the background equations, so \(\psi_k\) is auxiliary on-shell. This elimination is performed with the Weyl sector switched off. The Weyl sector is then restored separately: in the present calculation its contribution is evaluated perturbatively on the \(F_0=0\) light mode, for scalars by evaluating \(L_W\) on the constraint solution and eliminating \(\beta_k''\) with the leading equation, which changes \(n_s\) by less than \(10^{-5}\), and for tensors as described below. With \(F_0\neq0\), the local higher-derivative theory also contains an additional branch with an opposite-sign kinetic term. Under the complete-form-factor interpretation this branch is not quantized and carries no independent initial data; its classical contribution to \(\zeta_k\), once residual time reparametrizations cancel its constant superhorizon term, decays outside the horizon for \(F_0<0\) in the dS and quasi-dS analysis of Ref.~\cite{Hirano:companion} (Secs.~6.2 and A.2); it has not been re-evaluated on the exact background. Eliminating \(\psi_k\) at \(F_0=0\) gives \(L=\mathcal K\beta_k'^2+\mathcal M\beta_k^2\), with
\begin{align}
\mathcal K&=K_{\beta'\beta'}-\frac{K_{\beta'\psi}^2}{4K_{\psi\psi}},\\
\mathcal M&=K_{\beta\beta}-\frac{K_{\beta\psi}^2}{4K_{\psi\psi}}+\frac12\frac{d}{d\eta}\!\left(\frac{K_{\beta\psi}K_{\beta'\psi}}{2K_{\psi\psi}}\right),
\end{align}
in terms of the coefficients \(K_{XY}\) of the averaged Lagrangian. The coefficients involve up to five conformal derivatives of \(\bar a\), which are obtained algebraically from the background state using the trace equation, so no numerical differentiation is needed. The mode equation \((\mathcal K\beta_k')'=\mathcal M\beta_k\) is integrated from \(k/(aH)=80\), with Bunch--Davies initial conditions in WKB form for \(\sqrt{\mathcal K}\beta_k\), to \(k/(aH)=10^{-3}\), where \(\zeta_k=\beta_k/\epsilon_H\) is frozen; tilts are obtained by symmetric differences in \(\ln k\). For \(r_T\) we multiply the exact \(F_0=0\) tensor amplitude by the dS light-branch factor \(3bs_*/(3bs_*-2F_0(H_*))\), with \(s=4-a/(3b)+aH_0^2/(6bH^2)\), which is exact on dS. Order reduction of the Weyl term on the light branch provides a consistency check. It differs from the reduction that removes the anomaly-supported phase~\cite{Simon:1991bm,Parker:1993dk}: there the full semiclassical equations are reduced perturbatively about Einstein gravity, whereas here the \(a\)- and \(b\)-dependent background is kept exact and only the small Weyl correction to the perturbation equations is reduced.

The quadratic actions were derived independently in two ways: symbolically on a general \(\bar a(\eta)\), and sector by sector in Mathematica at fixed order in the displacement from dS. The reduced scalar coefficients and the tensor coefficients agree to all 16 digits at zeroth, first and second order in that displacement, and the second-order background is confirmed by the on-shell vanishing of the \(\psi_k'^2\) coefficient. The benchmark background and the Einstein-frame scalaron reference were also reproduced in a separate Mathematica implementation. Varying every numerical parameter changes \(n_s\) by at most \(3\times10^{-5}\), and \(A_s\) and \(r_T\) by a few parts in \(10^5\). These are numerical uncertainties. The Weyl correction to \(r_T\), \(2.7\%\) at the benchmark, carries a separate theoretical uncertainty. The dS light-branch normalization \(D_0=3bs-2F_0\) is exact on dS (Sec.~7.2 of Ref.~\cite{Hirano:companion}), and its quasi-dS corrections are controlled by \(\epsilon_{H*}\) and \(p_+\) (Sec.~7.4 of Ref.~\cite{Hirano:companion}). Relative to the Weyl correction, these corrections and the neglected terms of second order in \(F_0/(3bs)\) are of order \(\epsilon_{H*}\simeq0.008\), \(p_+\simeq0.002\), and \(2|F_0|/(3bs)\simeq0.03\), suggesting a relative uncertainty of order \(10^{-3}\) in \(r_T\), assuming no enhancement of the omitted terms. Order reduction on the exact background gives a consistent value, \(r_T=2.857\times10^{-3}\), against \(2.860\times10^{-3}\); because the reduced action is not perturbative deep inside the horizon, we use it only as a consistency check.

\section*{Appendix C: The scalaron limit}
As \(a\to0\) at fixed \(b\), Eq.~\eqref{eq:action} reduces to \(R+R^2\) gravity, which we compute independently in the Einstein frame. There it is a scalar with potential \(V=\tfrac34M_s^2M_{\rm Pl}^2(1-e^{-\sqrt{2/3}\,\phi/M_{\rm Pl}})^2\), with the Mukhanov--Sasaki equation for \(\zeta\) and the standard tensor equation. The pivot is matched physically: comoving \(k\) and conformal time are frame-independent, and \(k_*=a_JH_J\) is taken 60 Jordan-frame \(e\)-folds before the Jordan-frame end of acceleration, with \(a_J=a_E e^{-\phi/(\sqrt6M_{\rm Pl})}\). This gives \(n_s=0.96653\), \(r=3.2345\times10^{-3}\) and \(bA_s=201.0\). Our computation approaches these values linearly in \(p_+N_*\) (Table~\ref{tab:scalaronlimit}).
\begin{table}[h]
\caption{Approach to the scalaron limit: \(\mathcal R_n\equiv\Delta n_s/p_+\), \(\mathcal R_r\equiv(\Delta r_T/r^{\rm scal}_T)/(p_+N_*)\) and \(\mathcal R_A\equiv(\Delta A_s/A^{\rm scal}_s)/(p_+N_*)\), where \(\Delta\) denotes the anomaly-driven value minus the Einstein-frame \(R+R^2\) value, at \(N_*=60\), \(F_0=0\) and fixed \(b\).}
\label{tab:scalaronlimit}
\begin{ruledtabular}
\begin{tabular}{ccccc}
\(b/a\) & \(p_+N_*\) & \(\mathcal R_n\) & \(\mathcal R_r\) & \(\mathcal R_A\) \\
\hline
34.5 & 0.0966 & \(-1.02\) & \(-0.94\) & \(+1.07\) \\
100 & 0.0333 & \(-1.00\) & \(-0.97\) & \(+1.03\) \\
300 & 0.0111 & \(-1.00\) & \(-0.98\) & \(+1.03\) \\
1000 & 0.00333 & \(-0.9\pm0.1\) & \(-0.99\) & \(+1.06\) \\
\end{tabular}
\end{ruledtabular}
\end{table}
This is the origin of Eq.~\eqref{eq:deviations}, together with \(A_s/A_s^{\rm scal}\simeq1+p_+N_*\) at fixed \(b\). At large \(b/a\), where \(\ddot h\) becomes small, the background is integrated in the variables \(u=-\epsilon_H\) and \(w=\ddot h/h^3\) to preserve accuracy.

\section*{Appendix D: Weyl-sector stability along the trajectory}
The pair \(z_*,\bar z_*\) solves \(B+(z-2)F_W(z)/c=0\), with \(B=3bs/c\), on the physical sheet of \(F_W\), reached from real \(z\) below the cut by continuation in \(\sqrt{9/4-z}\). With the retarded prescription, the Green function of the pair contains both late-time exponents \(\Delta_\pm=\tfrac32\pm\sqrt{9/4-z_*}\), and it decays under Hubble friction if \({\rm Re}\,\Delta_->0\), i.e.\ \(|{\rm Im}\sqrt{z_*-9/4}|<3/2\). This defines \(\alpha_{\min}(B)\), which is \(2.5\) at the pivot curvature of the benchmark (\(B\simeq420\)) and grows slowly as \(H\) decreases. With the on-shell running, \(\alpha(H)=\alpha_0-\ln(H_0/H)\), where \(\alpha_0\equiv-F_0(H_0)/c\). The criterion is derived on dS and is controlled in the quasi-dS regime, \(\epsilon_H\lesssim0.1\), where it requires \(\alpha_0\gtrsim5\). We adopt \(\alpha_0=7\), which keeps \({\rm Re}\,\Delta_-\gtrsim0.5\) throughout that regime. The criterion then also holds formally up to \(\epsilon_H=1\), but it is not controlled there; through the exit we retain the perturbative light-branch prescription, which is an assumption rather than a demonstrated stability result. Its effect on the observables is small: \(r_T\) changes by about \(1\%\) between \(\alpha_0=5\) and \(7\), and \(n_s\) not at all. The analysis also assumes that the complex pair lies within the conformal window: at the benchmark it sits at \(|m_*|\simeq7H_0\) on dS, decreasing to about \(2H_0\) at the pivot, so \(\Lambda_{\rm EFT}\) must lie well above these scales. At the other end, a real pole below the continuum requires \(\alpha>\alpha_{\rm crit}\simeq4B+1+2\ln2\approx1.7\times10^3\). The window refers to the total coupling: a local \(\kappa_W\) shifts \(F_0\) and is admissible as long as \(\alpha\) stays within it. In the threshold-matching picture, \(\alpha\simeq\ln(H/m_{\rm gap})+\mathcal O(1)\), so \(\alpha_0\simeq7\) corresponds to \(m_{\rm gap}\sim10^{-3}H_0\). This is well below the curvature at the end of inflation, \(\simeq0.03H_0\), so the conformal sector still confines after inflation.


\end{document}